\documentclass[a4paper,12pt]{article}
\usepackage{amsthm}
\usepackage{amssymb}
\usepackage{color}
\usepackage{amsmath}
\usepackage{amsfonts}

\newtheorem{remark}{Remark}[section]

\newtheorem{lemma}{Lemma}[section]
\newtheorem{theorem}{Theorem}[section]

\newtheorem{corollary}{Corollary}[section]

\def\b1{\mbox{\boldmath $1$}}

\newenvironment{demo*}{\vspace{3mm}\noindent{\bf Proof.}}{\hfill $\Box$ \vspace{3mm}}

\begin{document}

\title{\bf \Large On the  optimal dividend problem  for a spectrally positive L\'evy process}
\author{\normalsize{\sc  Chuancun Yin},  {\sc Yuzhen Wen } and {\sc Yongxia Zhao}\\
%\footnote{Corresponding author: E-mail address: ccyin@mail.qfnu.edu.cn}\\
[3mm] {\normalsize\it  School of Mathematical Sciences, Qufu Normal University}\\
{\normalsize\it Shandong 273165, P.R.\ China} \\
 e-mail: ccyin@mail.qfnu.edu.cn} \maketitle
 \centerline{\large {\bf Abstract}}
\vskip0.01cm
In this paper we study the  optimal  dividend problem     for a company whose surplus process evolves as
 a  spectrally positive  L\'evy process before dividends are deducted. This model includes the dual model of the classical risk model and the dual model with diffusion as special cases. We assume that
dividends are paid to the shareholders according to an admissible strategy whose dividend rate
is bounded by a constant. The objective is to find a dividend policy so as to maximize the
expected discounted value of dividends which are paid to the shareholders until the company
is ruined. We show that the optimal dividend strategy is formed by a threshold strategy.\\

\noindent {\bf Keywords}: \, {Threshold strategy, Dual model, Optimal dividend strategy,
 Scale functions,  Spectrally positive  L\'evy process,
Stochastic control.\\

%\newpage
\normalsize

\baselineskip=16pt

%\noindent{\bf 1.~~Introduction}
\section{\normalsize INTRODUCTION}\label{intro}

Recently, dividend optimization problems for financial and insurance corporations have attracted extensive
attention. How should corporation pay  dividends to its shareholders? A possible goal is that the company tries to maximize the expectation of the discounted dividends until possible  ruin of the company.   In recent years, quite a few interesting papers deal with   the optimal dividend problem in the dual model. The dual model is an appropriate
model for a company driven by inventions or discoveries.  Other examples are commission-based businesses, such
as real estate agent offices or insurance  annuity business.

For example, Avanzi, Gerber and  Shiu (2007) considered the  model which is dual to the classical risk model, where the authors studied how the expectation of the discounted dividends until ruin can be calculated when gain distribution has an exponential distribution or mixtures of exponential distributions and show how the exact value of the optimal dividend barrier can be determined. Avanzi and Gerber (2008) examined  the same problem for the dual model  perturbed by diffusion. Moreover, they pointed out that ``the optimal dividend strategy in the dual model is  a barrier strategy. A direct proof that an optimal strategy is a barrier strategy is of some interest but has not been given to our knowledge; the proof in Bayraktar and Egami (2008) is for exponential gains only."  Yao,   Yang and Wang (2010) considered the optimal problem with dividend payments and issuance of equity in a dual risk model without a diffusion, assuming proportional transaction costs, they found optimal strategy which maximizes the expected present value of the dividends payout minus the discounted costs of issuing new equity before ruin. In addition, for exponentially distributed jump sizes, closed form solutions are obtained.
Dai,   Liu  and  Luan (2010, 2011) considered the same problem as in Yao et al. (2010) for a dual risk model with a diffusion with bounded gains and exponential gains, respectively.    Avanzi, Shen and Wong (2011) determined an explicit form for the value function in the dual model with diffusion when the gains distribution is a mixture of exponentials. They  showed that a  barrier  dividend strategy is also optimal and   conjectured that the optimal dividend strategy in the dual model with diffusion should be the barrier strategy, regardless of the gains distribution.   Recently,   Bayraktar, Kyprianou and Yamazaki (2013) using the fluctuation theory of  spectrally positive  L\'evy processes, show  the optimality of barrier strategies for all such L\'evy processes.

All of above-mentioned papers deal with  dual risk models with  barrier strategy. Such a strategy has a parameter $b>0$, the level of the barrier. Whenever the surplus exceeds the barrier, the excess is paid out immediately as a dividend. Barrier strategies often serve as candidates for the optimal strategy when the dividend
rate is unrestricted. However,   if a barrier strategy is applied, ultimate ruin of the company   is certain. In many circumstances this is not desirable. This consideration leads us to impose restriction on the dividend stream.  Ng (2009) considered the dual of the compound Poisson model under a threshold dividend strategy and derived a set of two integro-differential equations satisfied by the expected total discounted dividends until ruin and showed how the equations can be solved by using only one of the two integro-differential equations. The cases where profits follow an exponential or a mixture of exponential distributions are then solved and the discussion for the case of a general profit distribution follows by the use of Laplace transforms. He illustrated how the optimal threshold level that maximizes the expected total discounted dividends until ruin can be obtained. In this paper we  provide a uniform mathematical framework to analyze  the  optimal control   problem  with dividends payout  for a general spectrally positive  L\'evy process when the dividend rate is restricted.    This problem has been considered by
 Asmussen and Taksar (1997), Jeanblanc-Picqu\'e  and Shiryaev (1995) and  H$\phi$jgaard and Taksar (1999) in the diffusive case.

The rest of the paper is organized as follows. Section 2 presents the model and formulates the dividend
optimization problem. Section 3 discusses the threshold strategies.
Explicit expressions for the expected discounted value of dividend payments are obtained,
and in Section 4 we give the main results; it is shown that the optimal dividend
strategy is formed by a threshold strategy. This
strategy also  called the refraction strategy, prescribes paying no dividends when the net surplus of the
company is below an optimal barrier  $b^*$, and paying dividends at the fixed maximal rate $\alpha$ when net
surplus exceeds  $b^*$.

\section{\normalsize The model and the optimization problem}\label{math}

Let  $X = \{X_t\}_{t\ge 0}$ be a spectrally positive    L\'evy process with non-monotone paths on a
 filtered probability space $(\Omega,
{\cal F}, {\Bbb F}, P)$, where $\Bbb{F}=({\cal F}_t)$$_{t\ge 0}$ is
  generated by the process $X$ and  satisfies the usual conditions. The L\'evy triplet of $X$ is given by $(c, \sigma,\Pi)$, where  $\sigma\ge 0$ and $\Pi$ is a measure on $(0,\infty)$ satisfying
$$\int_{0}^{\infty}(1 \wedge x^2)\Pi(dx)<\infty.$$
Denote by $P_x$ for the  law of $X$  when $X_0=x$. Let $E_x$ be the
expectation  associated with   $P_x$. For short, we write $P$ and
$E$ when $X_0=0$.
  The  Laplace exponent of $X$ is given by
\begin{equation}
  \Psi (\theta)=\frac1{t}\log Ee^{-\theta X_t} =c\theta + \frac1 2\sigma^2\theta^2
+\int_{0}^{\infty}(e^{-\theta x}-1+\theta x\text{\bf
1}_{\{0<x<1\}})\Pi (dx), \label{math-eq1}
\end{equation}
where ${\text{\bf 1}}_A$ is the indicator function of a set $A$. In the sequel, we   assume that $-\Psi'(0+)=\Bbb{E}(X_1)>0$ which implies
the process $X$ drifts to $+\infty$. It is well known that if $\int_1^{\infty}y\Pi(dy)<\infty$, then $E(X_1)<\infty$, and $E(X_1)=-c+\int_1^{\infty}y\Pi(dy).$
Note that $X$ has paths of bounded variation if and only if
$$\sigma=0\;\;{\rm and}\;\; \int_{0}^{\infty}(1 \wedge x)\Pi(dx)<\infty.$$
In this case, the Laplace exponent (2.1)  can be written as
\begin{equation}
  \Psi (\theta) =c_0\theta + \frac1 2\sigma^2\theta^2
+\int_{0}^{\infty}(e^{-\theta x}-1)\Pi (dx), \label{math-eq2}
\end{equation}
with $c_0=c+\int_0^1 x\Pi(dx)$ the so-called drift of $X$.

For an arbitrary spectrally positive  L\'evy process, the  Laplace exponent  $\Psi$ is  strictly convex on $(0,\infty)$ and $\lim_{\theta\to\infty}\Psi(\theta)=\infty$. Thus there exists a function $\Phi:[0,\infty)\rightarrow [0,\infty)$
defined by $\Phi(q)=\sup\{\theta\ge 0|\Psi(\theta)=q\}$ such that
$\Psi(\Phi(q))=q,\ q\ge 0.$

For more details on spectrally positive  L\'evy processes, the reader is referred to  Bertoin (1996) and Kyprianou (2006).

Assume the canonical decomposition of $X$ is given by
\begin{equation}X_t=-ct+\sigma B_t+J_t, \; t\ge 0, \label{math-eq3}
\end{equation}
  where  $\{B_t, t\ge 0\}$ is a standard Wiener process,
$\{J_t, t\ge 0\}$ is a pure upward jump L\'evy process that is independent of  $\{B_t, t\ge 0\}$. In addition $J_0=0$.
 Note that the dual model with diffusion in Avanzi and Gerber (2008) corresponds to the case in which
$\Pi(dx)=\lambda F(dx)$, where $\lambda>0$ is the Poisson parameter and $F$ is the distribution of individual gains, and the rate of expenses is given by
$c_0=c+\int_0^1 x\Pi(dx).$ In particular, when $\sigma=0$, the model reduces to the so-called dual model in Avanzi, Gerber and Shiu (2007).

We now recall the definition of the $q-$scale function $W^{(q)}$. For  each $q\ge 0$ there exits a
continuous and increasing function $W^{(q)}:\Bbb{R}\rightarrow
[0,\infty)$, called the $q$-scale function defined in such a way
that $W^{(q)}(x) = 0$ for all $x < 0$ and on $[0,\infty)$ its
Laplace transform is given by
\begin{equation}
\int_0^{\infty}\text{e}^{-\theta
x}W^{(q)}(x)dx=\frac{1}{\psi(\theta)-q},\; \theta
>\Phi(q). \label{math-eq4}
\end{equation}
Closely related to $W^{(q)}$ is the function $Z^{(q)}$ given by
$$Z^{(q)}(x)=1+q\int_0^x W^{(q)}(y)dy,\ x\in \Bbb{R}.$$
 We will also use the following function
 $$\overline{Z}^{(q)}(x)=\int_0^x Z^{(q)}(z)dz, \ x\in \Bbb{R}.$$
Note that
$$Z^{(q)}(x)=1,\ \ \overline{Z}^{(q)}(x)=x,\ \ x\le 0.$$
The following facts about   the scale functions are taken from  Chan, Kyprianou and Savov (2011). If X has paths of bounded
variation then, for all $q\ge 0$, $W^{(q)}|_{(0,\infty)}\in
C^1(0,\infty)$ if and only if $\Pi$ has no atoms. In the case that
$X$ has paths of unbounded variation, it is known that, for all
$q\ge 0$, $W^{(q)}|_{(0,\infty)}\in C^1(0,\infty)$. Moreover if
$\sigma> 0$ then $C^1(0,\infty)$ may be replaced by $C^2(0,\infty)$.
Further, if the L\'evy measure has a density, then the scale
functions are always differentiable. It is well known that
 \begin{equation}
 W^{(q)}(0+)=\left\{
  \begin{array}{ll}\frac{1}{c_0}, & {\rm if} \ X\ {\rm has\ paths\ of\ bounded\ variation},\\
    0, &{\rm otherwise},
  \end{array}
  \right. \nonumber
\end{equation}
and
 \begin{equation}
 {W^{(q)}}'(0+)=\left\{
  \begin{array}{lll}\frac{2}{\sigma^2}, & {\rm if} \ \sigma\neq 0,\\
    \frac{q+\Pi(0,\infty)}{c_0^2}, & {\rm if} \ X\ {\rm is \ compound\ Poisson}\\
    \infty, & {\rm if} \ \sigma=0\ {\rm and} \ \Pi(0,\infty)=\infty.
  \end{array}
  \right. \nonumber
\end{equation}
In all
cases, if  $q>0$, then
$W^{(q)}(x)\sim e^{\Phi(q)x}/\Psi'(\Phi(q))$ as $x\to\infty$.

We assume that the surplus process of the company is modeled by (2.3) if no dividends are paid.
An admissible (dividend) strategy $\pi=\{L_t^{\pi}|t\ge 0\}$ is given by a nondecreasing, right-continuous and $\Bbb{F}$-adapted process starting at $0$.
 Let $U^{\pi}=\{U_t^{\pi}:t\ge 0\}$   be the company's surplus, net of dividend payments, at time t. Thus,
 $$U_t^{\pi}=X_t-L_t^{\pi},\ \ t\ge 0.$$
 In this article   we are interested
in the case that $\pi$ only admits absolutely continuous strategies such that
 \begin{equation}
 dL^{\pi}_t=l^{\pi}(t)dt,
 \end{equation}
and for $t\ge 0$, $l^{\pi}(t)$ satisfies
\begin{equation}
0\le l^{\pi}(t)\le \alpha,
\end{equation} where $\alpha$ is a ceiling rate.
We define the dividend-value function $V_{\pi}$ by
$$V_{\pi}(x)=E\left[\int_0^{\tau_{\pi}}e^{-q t}l^{\pi}(t)dt|U_0^{\pi}=x\right],$$
 where $q>0$  is an interest force for the calculation of the present value and $\tau_{\pi}$ is the time of ruin which is defined by
$$\tau_{\pi}=\inf\{t>0|U_t^{\pi}=0\}.$$
We denote by $\Xi$ the set of all the admissible dividend strategies. The objective is to   solve  the following stochastic control problem:
the maximal dividend-value function, which is defined as
\begin{equation}
V(x)=\sup_{\pi\in\Xi}V_{\pi}(x),
\end{equation}
and to find an optimal policy $\pi^*\in\Xi$ that satisfies $V(x)=V_{\pi^*}(x)$  for all $x\ge 0$.
In this paper,  we will prove that the optimal dividend strategy is formed by a threshold strategy with parameters $b^*$ (the definition
of $b^*$  is given by (4.4)) and $\alpha$: whenever the  controlled risk process is below $b^*$, no dividends are paid; however, when the
 controlled risk process is above this level, dividends are paid continuously at the maximal admissible rate $\alpha$.

\setcounter{equation}{0}
\section{ Threshold dividend strategies}\label{Thr}

In this section,  we assume that
the company pays dividends according to the following threshold  strategy
governed by parameters $b>0$ and $\alpha>0$ when risk process is modeled by $X$. Whenever the modified
surplus is below the threshold level $b$, no dividends are paid.
However,   when the surplus is above this threshold level, dividends
are paid continuously at a constant rate $\alpha$.   We define the modified   risk process
$U_b=\{U_b(t):t\ge 0\}$ in which $U_b(t)$ is  the solution to the stochastic differential equation given by
$$dU_b(t)=dX_t-\alpha  {\text{\bf 1}} _{\{U_b(t)>b\}}dt,\  t\ge 0.$$
  In the case of  $X$ is the dual of the compound Poisson model, the risk process  has been studied   by
Ng (2009)  in considerable detail.  For the general spectrally positive L\'evy process $X$, note that $-X$ is a  spectrally negative L\'evy process, the existence of  unique strong  solution to above equation follows directly from Kyprianou and Loeffen (2010) in which the so called refracted   L\'evy processes were
 established for spectrally negative L\'evy processes.

 Let $D_b$ denote the
present value of all dividends until time of ruin $T$. That is
$$D_b=\alpha\int_0^T e^{-q t} {\text{\bf 1}}_{\{U_b(t)>b\}}dt,$$  where
   $T=\inf\{t>0: U_b(t)=0\}$
with $T=\infty$ if $U_b(t)>0$ for all $t\ge 0$. Here $q>0$ is the discount rate.   Denote by   $V(x,b)$ the expected discounted
value of dividend payments, that is,
$$V(x,b)=E(D_b|U_b(0)=x).$$
Clearly, $0\le V(x,b)\le \frac{\alpha}{q}$ and
$\lim_{x\to\infty}V(x,b)=\frac{\alpha}{q}$.

Denote by ${\cal{A}}$ the extended generator of the process $X$, which acts on sufficiently smooth   functions $g$ defined by
\begin{equation}
{\cal{A}} g(x)=\frac{1} {2}\sigma^2 g''(x)-c g'(x)
+\int_{0}^{\infty}[g(x+y)-g(x)-g'(x)y\text{\bf
1}_{\{0<y<1\}}]\Pi(dy).\label{Thr-eq1}
\end{equation}

Throughout this paper a function $f:D \rightarrow (0,\infty)$ is called sufficiently smooth meaning
that it belongs to $C^1(D)$ if $X$ is of bounded variation, it belongs to $C^2(D)$ if $X$ has a
Gaussian exponent and it is twice continuously differentiable almost everywhere but is
not in $C^2(D)$ if $X$ is of unbounded variation and $\sigma=0$.

\begin {theorem} \label{thrm3-1}  Assume that, as a function of $x$,  $V(x,b)$ is  sufficiently smooth
 on $(0, b)\cup (b,\infty)$.
 Then $V(x,b)$ satisfies the following
integro-differential equations
\begin{equation}
{\cal A}V(x,b)=q V(x,b), \; 0<x<b,
\end{equation}
\begin{equation}
{\cal A}V(x,b)-\alpha V'(x,b)=q V(x,b)-\alpha, \; x>b,
\end{equation}
with initial condition $V(0,b)=0$ and continuity condition
\begin{equation}
V(b-,b)=V(b+,b)=V(b,b).
\end{equation}
In addition,
\begin{eqnarray}
&&(c_0+\alpha)V'(b+,b)-c_0 V'(b-,b)=\alpha,\ {\rm if}\   X\ {\rm is \ of\ bounded\ variation},\\
&&V'(b+,b)=V'(b-,b),\ {\rm if}\   X\ {\rm is \ of\ unbounded\ variation.}
 \end{eqnarray}
\end{theorem}
{\bf Proof}.\ If $\sigma>0$, then applying It\^{o}'s formula for semimartingales (cf. Klebaner (2008, P. 234) one has
\begin{eqnarray*}
&&E_x\left[e^{-q(t\wedge T)}V(U_b(t\wedge T),b)\right]=V(x,b)\\
&&+E_x \int_0^{t\wedge T}e^{-q s} [({\cal A}-q)V(U_b(s),b)-\alpha  {\text{\bf 1}}_{\{U_b(s)>b\}}V'(U_b(s),b)]ds.
\end{eqnarray*}
Letting $t\to\infty$ and note that  $V(0,b)=0$ (since $U_b$ always creeps downward and hence $U_b(T)=0$ almost surely given $T<\infty$) we have
$$V(x,b)=\alpha E_x \int_0^T e^{-q t} {\text{\bf 1}} _{\{U_b(t)>b\}}dt$$
if and only if
$$({\cal A}-q)V(x,b)-\alpha  {\text{\bf 1}}_{\{x>b\}}V'(x,b)=-\alpha {\text{\bf 1}}_{\{x>b\}}.$$
If $X$ is of bounded variation, we are allowed to use the change of variables (Theorem 31,
Protter, 1992); If $X$ is of unbounded variation and $\sigma = 0$, using the Meyer-Ito's formula
(Theorem 70, Protter, 1992) and stochastic integration by parts for semimartingales, we also get the result.
 The conditions (3.4) and (3.5) are true for the dual of the compound Poisson model (see Ng (2009)). For general case, we can prove (3.4)-(3.6) are also true by using  the approximation argument that   used in Shen et al. (2013) for spectrally negative L\'evy processes.
This ends the proof.

 Define the first passage times, with the convention $\inf
\emptyset=\infty$,
$$T^+_b=\inf\{t\ge 0: U_b(t)>b \},\;\;\;T_b^-=\inf\{t\ge 0: U_b(t)<b \}.$$
For $q\ge 0$, let
$$\Phi_1(q)=\sup\{\theta\ge 0|\Psi(\theta)+\alpha\theta=q\}.$$
The following result generalized the result of Ng (2009, Theorem 2) in which only the dual of the classical insurance risk model was considered.

\begin {theorem} \label{thrm3-2} For $x>b$, we have
\begin{equation}
V(x,b)=\frac{\alpha}{q}+\left(V(b,b)-\frac{\alpha}{q}\right)e^{-\Phi_1(q)(x-b)}.
\end{equation}
\end{theorem}
{\bf Proof}.\  By using the strong Markov property of $U_b$ at $T_b^-$ as in Yin et al. (2013), we have
 \begin{eqnarray*} V(x,b)&=&\frac{\alpha}{q}-\frac{\alpha}{q}E_x(e^{-q T_b^-}, T_b^-<\infty)\\
 &&+E_x(e^{-q T_b^-}V(U_b(T_b^-),b), T_b^-<\infty).
 \end{eqnarray*}
 The result (3.7) follows since $P_x(U_b(T_b^-)=b, T_b^-<\infty)=1$ and
 $$E_x(e^{-q T_b^-}, T_b^-<\infty)=\exp(-\Phi_1(q)(x-b)).$$
 The last formula is obtained from the corresponding result for a spectrally negative L\'evy process (cf. Bertoin (1996, Theorem 1, P.189)). This completes the proof.

\begin {theorem} \label{thrm3-3} Let  $X = \{X_t\}_{t\ge 0}$ be a spectrally positive    L\'evy process, we assume that the L\'evy measure  $\Pi$ has no atoms in the case
that $X$ has paths of bounded variation. Then
for $0<x<b$, we have
\begin{eqnarray}
V(x,b)&=&\frac{\alpha}{q}Z^{(q)}(b-x)\nonumber\\
&+&\left(V(b,b)-\frac{\alpha}{q}\right)e^{\Phi_1(q)(b-x)}\left(1+\alpha\Phi_1(q)\int_0^{b-x}W^{(q)}(z)
e^{-\Phi_1(q)z}dz\right),
\end{eqnarray}
where
$$V(b,b)=\frac{\alpha}{q}\left(1-\frac{Z^{(q)}(b)e^{-\Phi_1(q)b}}{1+\alpha\Phi_1(q)\int_0^{b}W^{(q)}(z)
e^{-\Phi_1(q)z}dz}\right).$$
\end{theorem}
 {\bf  Proof}.\ By the law of total probability and the strong Markov property,  for $0<x<b$, we have
$$V(x,b)=E_x(e^{-qT_b^+}V(X_{T_b^+}, b),  T_b^+<T).$$
This, together with (3.7),  yields
\begin{eqnarray}
 V(x,b)&=&\frac{\alpha}{q} E_x(e^{-qT_b^+},  T_b^+<T)\nonumber\\
 &&+\left(V(b,b)-\frac{\alpha}{q}\right)E_x(e^{-qT_b^{+}-\Phi_1(q)(X_{T_b^+}-b)},  T_b^+<T).
\end{eqnarray}
From (2.3) in Yin and Wen (2013), we have
\begin{eqnarray}
E_x (e^{-q T^+_b}, T_b^+<T_0^-)=Z^{(q)}(b-x)-W^{(q)}(b-x)\frac{Z^{(q)}(b)}{W^{(q)}(b)}.
\end{eqnarray}
Next, we compute the second expectation in (3.9).
By using the following formula, which  can be found in
Kuznetsov, Kyprianiu and Pardo (2012, P. 1117) or Kadankov and Kadankova (2005, (12)):
\begin{eqnarray}
E_x\left(e^{-qT_b^+}f(X_{T_b^+})\right)&=&E_x\left(e^{-qT_b^+}f(X_{T_b^+}), T_b^+<T_0^-\right)\nonumber\\
&&+E_x\left(e^{-qT_0^-}E_{X_{T_0^-}}\left(e^{-qT_b^+}f(X_{T_b^+})\right), T_b^+>T_0^-\right),\nonumber
\end{eqnarray}
and note that $P_x(X_{\tau_0^-}=0)=1$ and $P_x(T_0^-=T)=1$, we have
\begin{eqnarray}
E_x\left(e^{-qT_b^+-\Phi_1(q)X_{T_b^+}}, T_b^+<T_0^-\right)&=&-E_x\left(e^{-qT_0^-}, T_b^+>T_0^-\right) E_0\left(e^{-qT_b^+-\Phi_1(q)X_{T_b^+}}\right)\nonumber\\
&&+E_x\left(e^{-qT_b^+-\Phi_1(q)X_{T_b^+}}\right).
 \end{eqnarray}
From (2.2) in Yin and Wen (2013), we have
\begin{eqnarray}
E_x\left(e^{-qT_0^-}, T_b^+>T_0^-\right)=\frac{W^{(q)}(b-x)}{W^{(q)}(b)}.
\end{eqnarray}
Consider the dual process $-X$ and by virtue of (58) in Kuznetsov, Kyprianiu and Rivero (2012, P.122), we obtain
\begin{eqnarray}
E_x\left(e^{-qT_b^+-\Phi_1(q)X_{T_b^+}}\right)&=&e^{-\phi_1(q)x}\left(1+\alpha\Phi_1(q)\int_0^{b-x}W^{(q)}(z)
e^{-\Phi_1(q)z}dz\right)\nonumber\\
&&-\frac{\alpha \Phi_1(q)}{\Phi(q)-\Phi_1(q)}e^{-\phi_1(q)b}W^{(q)}(b-x).
 \end{eqnarray}
Substituting (3.10)-(3.13) into (3.9), we get
\begin{eqnarray}
V(x,b)&=&\frac{\alpha}{q}\left(Z^{(q)}(b-x)-W^{(q)}(b-x)\frac{Z^{(q)}(b)}{W^{(q)}(b)}\right)\nonumber\\
&&+\left(V(b,b)-\frac{\alpha}{q}\right)\left\{e^{\Phi_1(q)(b-x)}\left(1+\alpha\Phi_1(q)\int_0^{b-x}W^{(q)}(z)
e^{-\Phi_1(q)z}dz\right)\right.\nonumber\\
&&\left.-\frac{W^{(q)}(b-x)}{W^{(q)}(b)}e^{\Phi_1(q)b}\left(1+\alpha\Phi_1(q)\int_0^{b}W^{(q)}(z)
e^{-\Phi_1(q)z}dz\right)\right\},\nonumber
\end{eqnarray}
where the constant $V(b,b)$ can be determined by conditions (3.5) and (3.6).
 Since
\begin{eqnarray}
\frac{\alpha}{q}Z^{(q)}(b)
+\left(V(b,b)-\frac{\alpha}{q}\right) e^{\Phi_1(q)b}\left(1+\alpha\Phi_1(q)\int_0^{b}W^{(q)}(z)
e^{-\Phi_1(q)z}dz\right)=0,\nonumber
\end{eqnarray}
 the result (3.8) follows.

{\bf  Second proof of Theorem 3.3}.\  From (3.7) and (3.9) we see that $V$ has the assumed smoothness in Theorem 3.1. We first assume that $\lambda:=\int_0^{\infty}\Pi(dy)<\infty$.
Set
$c_0=c+\int_0^1 y\Pi(dy)$ and $f(y)dy=\frac{\Pi(dy)}{\lambda}$,
 then $f$ is a probability density on  $(0,\infty)$.
In this case, in view of (3.7), the integro-differential
equation (3.2) can be written as
\begin{eqnarray}
\frac12\sigma^2 V''(x,b)-c_0 V'(x,b)&=&-\lambda\left[V(b,b)-\frac{\alpha}{q}\right]\int_{b-x}^{\infty}\exp(-\Phi_1(q)(x+y-b))f(y)dy\nonumber\\
&&-\lambda\int_0^{b-x}V(x+y,b)f(y)dy-\frac{\lambda\alpha}{q}(1-F(b-x))\nonumber\\
&&+(\lambda+q)V(x,b), \ 0<x<b,
\end{eqnarray}
where $F$ is the distribution function of $f$.
Replace the variable $x$ by $z=b-x$, and define $W$ by $W(z,b)=V(b-z,b), 0<z<b$. The (3.14) becomes
 \begin{eqnarray}
\frac12\sigma^2 W''(z,b)+c_0 W'(z,b)&=&-\lambda\left[W(0,b)-\frac{\alpha}{q}\right]\int_{z}^{\infty}\exp(-\Phi_1(q)(y-z))f(y)dy\nonumber\\
&&-\lambda\int_0^{z}W(y,b)f(z-y)dy-\frac{\lambda\alpha}{q}(1-F(z))\nonumber\\
&&+(\lambda+q)W(z,b), \ 0<z<b,
\end{eqnarray}
with  initial condition $W(0,b)=V(b,b)$ and boundary condition $W(b,b)=0$.
 We extend the definition of $W$ by  (3.15) to $z\ge 0$ and denote the resulting function by $w$. Then we have
\begin{eqnarray}
\frac12\sigma^2 w''(z)+c_0 w'(z)&=&-\lambda\left[w(0)-\frac{\alpha}{q}\right]\int_{z}^{\infty}\exp(-\Phi_1(q)(y-z))f(y)dy\nonumber\\
&&-\lambda\int_0^{z}w(y)f(z-y)dy-\frac{\lambda\alpha}{q}(1-F(z))\nonumber\\
&&+(\lambda+q)w(z), \ z\ge 0,
\end{eqnarray}
with $w(0)=V(b,b)$ and $w(b)=0$.

For a function $g$, denoted by $\hat{g}$  the Laplace transform of $g$, i.e.
$\hat{g}(\xi)=\int_0^{\infty}e^{-\xi y}g(y)dy$.
Then the Laplace transform $\hat{w}$ for $w$ can be easily determined from Eq. (3.16) as
\begin{equation}
\hat{w}(\xi)=\frac{\frac12\sigma^2(w'(0)+\xi w(0))+c_0 w(0)+\frac{\lambda\alpha}{q\xi}(\hat{f}(\xi)-1)-\frac{\lambda}{\xi-\Phi_1(q)}(w(0)-\frac{\alpha}{q})
(\hat{f}(\Phi_1(q))-\hat{f}(\xi))}
{\frac12\sigma^2\xi^2+c_0\xi-(\lambda+q)+\lambda\hat{f}(\xi)}.
\end{equation}
Note that
$$\int_0^{\infty}e^{-x\xi}W^{(q)}(x)dx=\frac{1}{\frac12\sigma^2\xi^2+c_0\xi-(\lambda+q)+\lambda\hat{f}(\xi)},$$
$$\int_0^{\infty}e^{-x\xi}dx\int_0^x W^{(q)}(y)dy=\frac{1}{\xi(\frac12\sigma^2\xi^2+c_0\xi-(\lambda+q)+\lambda\hat{f}(\xi))},$$
$$\int_0^{\infty}e^{-x\xi}dx\int_0^x dy\int_0^y W^{(q)}(z)dz=\frac{1}{\xi^2(\frac12\sigma^2\xi^2+c_0\xi-(\lambda+q)+\lambda\hat{f}(\xi))}.$$
Now inverting (3.17) gives
\begin{eqnarray}
w(z)&=&\frac12w'(0)\sigma^2W^{(q)}(z)+\frac12\sigma^2 w(0)\int_0^z W^{(q)}(z-y)\delta_0'(y)dy\nonumber\\
&&+c_0 w(0) W^{(q)}(z)+\lambda w(0)\left(W^{(q)}*(f-\hat{f}(\Phi_1(q))\delta_0)*l\right)(z)\nonumber\\
&&+\frac{\lambda \alpha}{q}\left(\Phi_1(q)\frac{Z^{(q)}-1}{q}*(\delta_0-f)*l\right)(z)\nonumber\\
&&+\frac{\lambda \alpha}{q}\hat{f}(\Phi_1(q))-1)(W^{(q)}*l)(z),
\end{eqnarray}
where $\delta_0$ is the  Dirac delta function at $0$, $h_1*h_2$ stands for convolution of $h_1$ and $h_2$ and $l(z)=\exp(\Phi_1(q)z).$
After some tedious calculations, we get
$$\int_0^z W^{(q)}(z-y)\delta_0'(y)dy={W^{(q)}}'(z),$$
\begin{eqnarray}
\lambda\left(W^{(q)}*(f-\hat{f}(\Phi_1(q))\delta_0)*l\right)(z)&=&\alpha\Phi_1(q)\int_0^z W^{(q)}(z-y)e^{\Phi_1(q)y}dy\nonumber\\
&&+e^{\Phi_1(q)z}-\frac12\sigma^2{W^{(q)}}'(z)\nonumber\\
&&-(\frac12\sigma^2\Phi_1(q)+c_0)W^{(q)}(z), \nonumber
\end{eqnarray}
\begin{eqnarray}
\lambda((Z^{(q)}-1)*(\delta_0-f)*l)(z)&=&\left(\frac12\sigma^2 q\Phi_1(q)+c_0 q\right)\int_0^z W^{(q)}(z-y)e^{\Phi_1(q)y}dy\nonumber\\
&&+\frac12\sigma^2 q W^{(q)}(z)-q
\int_0^{z}Z^{(q)}(z-y)
e^{\Phi_1(q)y}dy.\nonumber
\end{eqnarray}
Substituting the three expressions above into (3.18)
we arrive at
\begin{eqnarray}
w(z)&=&\frac12\sigma^2 w'(0) W^{(q)}(z)\nonumber\\
&+&w(0)\left(e^{\Phi_1(q)z}-\frac12\sigma^2 \Phi_1(q) W^{(q)}(z)\right.\nonumber\\
&&\left.+\alpha \Phi_1(q)e^{\Phi_1(q)z}\int_0^{z}W^{(q)}(y)
e^{-\Phi_1(q)y}dy\right)\nonumber\\
&+& \frac{\alpha}{q}\left(\frac12\sigma^2 \Phi_1(q) W^{(q)}(z)-e^{\Phi_1(q)z}\Phi_1(q)
\int_0^{z}Z^{(q)}(y)
e^{-\Phi_1(q)y}dy\right)\nonumber\\
&+&\frac{\alpha}{q}e^{\Phi_1(q)z}\left(q-\alpha\Phi_1(q)\right)\int_0^{z}W^{(q)}(y)
e^{-\Phi_1(q)y}dy.\nonumber
\end{eqnarray}
Note that
$$\frac12\sigma^2 w'(0) W^{(q)}(z)-\frac12w(0)\sigma^2 \Phi_1(q) W^{(q)}(z)+\frac{\alpha}{q}\frac12\sigma^2 \Phi_1(q) W^{(q)}(z)=0,$$
and
\begin{eqnarray}
-e^{\Phi_1(q)z}\Phi_1(q)\int_0^{z}Z^{(q)}(y)e^{-\Phi_1(q)y}dy
&+&e^{\Phi_1(q)z}\left(q-\alpha\Phi_1(q)\right)\int_0^{z}W^{(q)}(y)
e^{-\Phi_1(q)y}dy\nonumber\\
&=&Z^{(q)}(z)-e^{\Phi_1(q)(z)}\left(1+\alpha\Phi_1(q)\int_0^{z}W^{(q)}(z)
e^{-\Phi_1(q)z}dz\right).\nonumber
\end{eqnarray}
Thus we have the simpler expression
\begin{eqnarray}
w(z)=\frac{\alpha}{q}Z^{(q)}(z)
+\left(w(0)-\frac{\alpha}{q}\right)e^{\Phi_1(q)(z)}\left(1+\alpha\Phi_1(q)\int_0^{z}W^{(q)}(z)
e^{-\Phi_1(q)z}dz\right),
\end{eqnarray}

and the result (3.8) follows since $V(x,b)=w(b-x)$ and $w(0)=V(b,b)$.

Now, we assume that
$\lambda:=\int_0^{\infty}\Pi(dy)=\infty$.  Let $\Pi_n$ be measures on  $(0,\infty)$ defined by
$$\Pi_n(dx)=\Pi(dx){\bf 1}_{\{(\frac{1}{n},\infty)\}}(x),\ n\ge 1.$$
Then we have
$$\lambda_n:=\int_0^{\infty}\Pi_n(dx)\le n^2\int_{\frac{1}{n}}^1 x^2\Pi(dx) +\int_1^{\infty}(1\wedge x^2)\Pi(dx)<\infty.$$
Set
$c_n=c+\int_0^1 y\Pi_n(dy)$ and $f_n(y)dy=\frac{\Pi_n(dy)}{\lambda_n}$,
 then for each $n\ge 1$, $f_n$ is a probability density on  $(0,\infty)$.
Similar to  (3.14) we consider the following integro-differential
equation
\begin{eqnarray}
\frac12\sigma^2 {V_n}''(x,b)-c_n {V_n}'(x,b)&=&-\lambda\left[V_n(b,b)-\frac{\alpha}{q}\right]\int_{b-x}^{\infty}\exp(-\Theta_n(q)(x+y-b))f_n(y)dy\nonumber\\
&&-\lambda\int_0^{b-x}V_n(x+y,b)f_n(y)dy-\frac{\lambda\alpha}{q}(1-F_n(b-x))\nonumber\\
&&+(\lambda+q)V_n(x,b), \ 0<x<b,\nonumber
\end{eqnarray}
where $\Theta_n(q)=\sup\{\theta\ge 0|\Psi_n(\theta)+\alpha\theta=q\}.$
Here,
\begin{equation}
  \Psi_n (\theta) =c_n\theta + \frac1 2\sigma^2\theta^2
+\int_{0}^{\infty}(e^{-\theta x}-1+\theta x\text{\bf
1}_{\{|x|<1\}})\Pi_n (dx).\nonumber
\end{equation}
Repeating the same argument as the case that $\lambda:=\int_0^{\infty}\Pi(dy)<\infty$, we obtain
 \begin{eqnarray}
V_n(x,b)&=&\frac{\alpha}{q}Z_n^{(q)}(b-x)\nonumber\\
&+&(V_n(b,b)-\frac{\alpha}{q})e^{\Theta_n(q)(b-x)}\left(1+\alpha\Theta_n(q)\int_0^{b-x}W_n^{(q)}(z)
e^{-\Theta_n(q)z}dz\right),\nonumber
\end{eqnarray}
where $W^{(q)}_n$ and  $Z^{(q)}_n$ are scale functions corresponding to the process $X_n$ with Laplace exponent $\Psi_n$.
Since $\lim_{n\to\infty}\Psi_n(\theta)=\Psi(\theta)$, then $X_n\to X$ weakly as $n\to\infty$.   Thus $\lim_{n\to\infty}V_n=V$,   $\lim_{n\to\infty}W^{(q)}_n=W^{(q)}$,
$\lim_{n\to\infty}Z^{(q)}_n=Z^{(q)}$ and $\lim_{n\to\infty}\Theta_n=\Phi_1$. Consequently, (3.8) still holds for this case. The constant $V(b,b)$ can be determined by conditions (3.5) and (3.6).  This ends the proof of Theorem 3.3.
\begin{remark}  For $\sigma\ge 0$,  from the graph of
$$\Psi(\Phi_1(q))+\alpha \Phi_1(q)=q$$
one can verify that $\Phi_1(q)\to 0$ when $\alpha\to\infty$.
After some simple calculations we get
$\lim_{\alpha\to\infty}\alpha \Phi_1(q)=q$,
$$\lim_{\alpha\to\infty}\alpha(q-\alpha\Phi_1(q))=q\Psi'(0+),$$
and
$$\lim_{\alpha\to\infty}V(b,b)=\frac{\overline{Z}^{(q)}(b)}{Z^{(q)}(b)}+\frac{\Psi'(0+)}{q Z^{(q)}(b)}-\frac{\Psi'(0+)}{q}.$$
As a result, for $0<x<b$, we arrive at
$$\lim_{\alpha\to\infty}V(x,b)=\frac{\overline{Z}^{(q)}(b)}{Z^{(q)}(b)}Z^{(q)}(b-x)-\overline{Z}^{(q)}(b-x)
+\frac{\Psi'(0+)}{q}\left(\frac{Z^{(q)}(b-x)}{Z^{(q)}(b)}-1\right),
$$
 which is the expected discounted
value of dividend payments for the barrier strategy. See Lemma 2.1 of  Bayraktar, Kyprianou and Yamazaki (2013).
\end{remark}

\setcounter{equation}{0}
\section{\normalsize  Optimal dividend strategy}\label{opt}
In this section we will focus on verifying the optimality of the threshold strategy with $b^*$.
\begin{lemma} (Verification lemma)\ Suppose that $\hat{\pi}$ is an admissible dividend strategy such that
$V_{\hat{\pi}}$ is smooth on $(0,\infty)$ and for all $x>0$
\begin{equation}
{\cal{A}} V_{\hat{\pi}}(x)-qV_{\hat{\pi}}(x)+  \sup_{0\le r\le \alpha}\{r(1-V_{\hat{\pi}}'(x))\}\le 0
 \end{equation}
with $V_{\hat{\pi}}(0)=0$,
 where ${\cal{A}}$ is  the extended generator of the process $X$,  which is defined by (3.1).
 Then $V_{\hat{\pi}}(x)=V(x)$ for $x\ge 0$ and hence $\hat{\pi}$ is an optimal strategy.
\end{lemma}
{\bf Proof}.\ Kyprianou, Loeffen and P\'erez (2012) obtained the related result for spectrally negative  L\'evy processes. The proof here is similar to the proof of Lemma 5 in their paper and we omit it.

\begin{theorem} For any  spectrally positive  L\'evy process,
 consider the stochastic control problem (2.7). Let $\Xi$ be the
class of admissible dividend strategies satisfying (2.5)-(2.6). Suppose that $\Phi_1(q)\frac{\alpha}{q}>1$.
 Then we have $V(x)=V(x,b^*)$  and   the
threshold strategy with threshold level $b^*$ is the optimal dividend strategy over $\Xi$,
where
\begin{equation}
 V(x,b^*)=\left\{
  \begin{array}{ll}\frac{\alpha}{q}-\frac{1}{\Phi_1(q)} e^{-\Phi_1(q)(x-b^*)}, & {\rm if} \ x>b^*,\\
   -\frac{\alpha\int_0^{b^*-x} W^{(q)}(z) e^{-\Phi_1(q)z}dz}{e^{-\Phi_1(q)(b^*-x)}}+\frac{\alpha}{q}Z^{(q)}(b^*-x)-\frac{e^{\Phi_1(q)(b^*-x)}}{\Phi_1(q)}, &{\rm if} \ 0<x<b^*,
  \end{array}
  \right.
\end{equation}
and $b^*$ is determined by
$$V(b^*,b^*)=\frac{\alpha}{q}-\frac{1}{\Phi_1(q)}.$$
\end{theorem}

\begin{corollary}
For the dual of the compound Poisson  model  perturbed by Brownian motion,  the optimal strategy with bounded rate of
dividend payment is formed by a threshold one
regardless of the    gains  distribution.
\end{corollary}

\begin{corollary}
 For the dual of the compound Poisson  model,  the optimal strategy with bounded rate of
dividend payment is formed by a threshold one provided
  the    gains  distribution is continuous on $(0,\infty)$.
\end{corollary}

{\bf Proof of Theorem 4.1}.\   It follows  from (3.7) that
\begin{equation}
V'(x,b)= -\Phi_1(q)\left(V(b,b)-\frac{\alpha}{q}\right)e^{-\Phi_1(q)(x-b)},\ x>b.
\end{equation}

If $\Phi_1(q)\frac{\alpha}{q}\le 1$, then  $V'(x,0)\le 1$ since $V(0,0)=0$. Thus $b^*=0$.
Now, suppose that $\Phi_1(q)\frac{\alpha}{q}>1$, we get  the condition for $b^*$:
\begin{equation}
 -\Phi_1(q)\left(V(b^*,b^*)-\frac{\alpha}{q}\right)=1. \nonumber
\end{equation}
Or, equivalently, $b^*$ is the solution of the equation
\begin{equation}
V(b^*,b^*)=\frac{\alpha}{q}-\frac{1}{\Phi_1(q)}.
\end{equation}
From (3.7) and (3.8)  we get
\begin{equation}
 V(x,b^*)=\left\{
  \begin{array}{ll}\frac{\alpha}{q}-\frac{1}{\Phi_1(q)} e^{-\Phi_1(q)(x-b^*)}, & {\rm if} \ x>b^*,\\
   -\frac{\alpha\int_0^{b^*-x} W^{(q)}(z) e^{-\Phi_1(q)z}dz}{e^{-\Phi_1(q)(b^*-x)}}+\frac{\alpha}{q}Z^{(q)}(b^*-x)-\frac{e^{\Phi_1(q)(b^*-x)}}{\Phi_1(q)}, &{\rm if} \ 0<x<b^*.
  \end{array}
  \right.
\end{equation}
Taking derivative with respect to $x$ in the both sides of the above relation leads to
\begin{equation}
 V'(x,b^*)=\left\{
  \begin{array}{ll} e^{-\Phi_1(q)(x-b^*)}, & {\rm if} \ x>b^*,\\
    e^{\Phi_1(q)(b^*-x)}\left(1+\alpha \Phi_1(q)\int_0^{b^*-x} W^{(q)}(z) e^{-\Phi_1(q)z}dz\right), &{\rm if} \ 0<x<b^*.
  \end{array}
  \right.
\end{equation}
It follows that $V'(x,b^*)<1$ when $x>b^*$.
Further, for  $0<x<b^*$,
\begin{eqnarray}
V''(x,b^*)&=&-\Phi_1(q)e^{\Phi_1(q)(b^*-x)}\left(1+\alpha \Phi_1(q)\int_0^{b^*-x} W^{(q)}(z) e^{-\Phi_1(q)z}dz\right)\nonumber\\
&-&\alpha \Phi_1(q) W^{(q)}(b^*-x)<0.
\end{eqnarray}
Thus for $0<x<b^*$, $V'(x,b^*)>V'(b^*,b^*)=1.$ Taking into account of  Eqs (3.2) and (3.3) we see that $V(x,b^*)$ satisfies (4.1) and the result follows.

\begin{remark} From (4.6) and (4.7) we  find that $V'(x,b^*)$ is continuous on $(0,\infty)$,
 $V''(b^*-,b^*)= -\Phi_1(q)-\alpha \Phi_1(q)  W^{(q)}(0)$ and
 $V''(b^*+,b^*)= -\Phi_1(q)$. So that if  $X$ has paths of bounded variation, then
 $V''(b^*-,b^*)\neq V''(b^*+,b^*)$ and, if   $X$ has paths of unbounded variation, then $V''(b^*-,b^*)=V''(b^*+,b^*)$.
\end{remark}

\begin{remark} It is interesting to note that  the optimal
strategy with bounded rate of dividend payment is formed by a  threshold
strategy for spectrally positive  L\'evy process regardless of the L\'evy measure. However, for the spectrally negative  L\'evy
model $-X$, it was shown in Kyprianou, Loeffen and P\'erez (2012)  that  the optimal
strategy   is formed by a  threshold
strategy   under   condition that the L\'evy measure  of $-X$ has a completely monotone density.
\end{remark}

\vskip0.3cm

\noindent{\bf Acknowledgements}\; The authors cordially thanks the Editor Prof. David Dickson
 and anonymous referee
 for many constructive suggestions and insightful  comments on the previous version of this paper.
The research was supported by the National Natural
Science Foundation of China (No.11171179) and the Research Fund for
the Doctoral Program of Higher Education of China (No. 20133705110002) and the Program for Scientific Research Innovation Team in Colleges and Universities of Shandong Province.

\end{document}